\documentclass[twocolumn,showpacs,preprintnumbers,prl,fleqn,floatfix]{revtex4}

\usepackage{graphicx}
\usepackage{dcolumn}
\usepackage{bm}
\usepackage{amsmath, amssymb}

\begin{document}

\title{Disorder-induced stabilization of the quantum Hall ferromagnet}

\author{B. A. Piot$^{1}$}

\altaffiliation{B.A. Piot and W. Desrat contributed equally to
this work.}

\author{W. Desrat$^{2}$}

\altaffiliation{B.A. Piot and W. Desrat contributed equally to
this work.}

\author{D. K. Maude$^{3}$}

\affiliation{$^{1}$ Laboratoire National des Champs Magn\'etiques
Intenses, LNCMI-CNRS-UJF-UPS-INSA-EMFL, F-38042 Grenoble, France}
\affiliation{$^{2}$Universit\'{e} Montpellier 2 and CNRS,
Laboratoire Charles Coulomb UMR 5221, F-34095, Montpellier,
France} \affiliation{$^{3}$ Laboratoire National des Champs
Magn\'etiques Intenses, LNCMI-CNRS-UJF-UPS-INSA-EMFL, F-31400
Toulouse, France}

\author{D. Kazazis$^{4}$, A. Cavanna$^{4}$, and U. Gennser$^{4}$}

\affiliation{$^{4}$Laboratoire de Photonique et de Nanostructures,
CNRS, Route de Nozay, 91460 Marcoussis, France}

\date{\today}

\begin{abstract}

We report on an absolute measurement of the electronic spin
polarization of the $\nu=1$ integer quantum Hall state. The spin
polarization is extracted in the vicinity of $\nu=1$ (including at
exactly $\nu=1$) via resistive NMR experiments performed at
different magnetic fields (electron densities), and Zeeman energy
configurations. At the lowest magnetic fields, the polarization is
found to be complete in a narrow region around $\nu=1$. Increasing
the magnetic field (electron density) induces a significant
depolarization of the system, which we attribute to a transition
between the quantum Hall ferromagnet and the Skyrmion glass phase
theoretically expected as the ratio between Coulomb interactions
and disorder is increased. These observations account for the
fragility of the polarization previously observed in high mobility
2D electron gas, and experimentally demonstrate the existence of
an optimal amount of disorder to stabilize the ferromagnetic
state.
\end{abstract}
 \maketitle

Two dimensional electron gases (2DEGs) under magnetic field
constitute a unique playground to study many-body effects. In a
strong enough perpendicular magnetic field, 2D electrons are
eventually confined to a single energy level, the lowest ``Landau
level'', in which Coulomb interactions can generate a series of
collective ground states. When the lowest Landau level is fully
occupied, which corresponds to a filling factor $\nu$ equal to
one, exchange interactions stabilize a ferromagnetic ground state
with long range order known as the ``$\nu=1$ quantum Hall
ferromagnet'' (QHF). The lowest energy excitations in this ground
state are generally not single spin flips but peculiar textures
involving the reversal of several spins, known as Skyrmions
\cite{Sondhi1993,Fertig1994,Schmeller1995,Maude1996}. Slightly
away from the exact $\nu=1$, charges can be added/removed from the
system by forming Skyrmions, which leads to a fast spin
depolarization evidenced e.g. by nuclear magnetic resonance (NMR)
``Knight shift'' measurements \cite{Barrett1995}. While early NMR
measurements gave substantial information on the Skyrmions
formation around $\nu=1$, an accurate reference for full spin
polarization is needed to appreciate the degree of polarization in
a more quantitative way. Subsequent ``absolute'' measurements of
the spin polarization \cite{Aifer1996,Zhitomirsky2004} have
actually revealed an incomplete polarization even at exact filling
factor $\nu=1$ where Skyrmions are a priori not expected. More
recently, very sharp depolarizations have been observed for small
filling factor deviations from $\nu=1$ and/or non
zero-temperatures \cite{Plochocka2009}. The origin of this
fragility of the spin polarization of the 2DEG around $\nu=1$ and
the condition for the stability of the QHF is the central focus of
our present work.

In this work, we present resistive measurements of the NMR Knight
shift providing us with an absolute determination of the $\nu=1$
QHF spin polarization at milliKelvin temperatures. Magnetic
field-dependent measurements performed by varying the electron
density of the 2DEG enable us to probe the spin polarization as a
function of the ratio  $\gamma_{int}$ between Coulomb interactions
and disorder. For the lowest explored values of $\gamma_{int}$,
where disorder is significant, the QHF is stabilized in the close
vicinity of $\nu=1$ where the polarization $P$ tends to 1.
However, as $\gamma_{int}$ is increased, a depolarization of the
system is observed. We attribute our observation to the
theoretically predicted phase transition between the QHF and the
quantum Hall Skyrmion Glass (QHSG), as interactions are increased
with respect to disorder. This demonstrates the existence of an
``optimal'' amount of disorder to stabilize the QHF, and in turns
accounts for the fragility of the spin polarization in high
mobility (high $\gamma_{int}$) samples reported in the literature.
The effect of the Zeeman energy on the polarization of the
skyrmion phase is finally examined via angular dependent
measurements, and is explained by the theoretically expected
changes in the skyrmion size.

The studied sample is a GaAs 2DEG in which the electron density
can be continuously tuned by the application of a top gate
voltage. The characteristics of the sample are summarized in table
\ref{table1}.
\begin{table}[h]
\begin{center}
\begin{tabular}{ccccccc}
\hline\hline $n$ ($cm^{-2}$) & $\mu$ ($10^{6}cm^{2}/Vs$) & $B_{F}$(T) & $B_{tot}$(T)& $\gamma_{int}$ & $\eta$ \\
\hline

 1.2 & 0.30 & 5   & 5 & 3.93 & 0.0123 & \\
 1.73 & 0.44& 7.2  & 7.2 & 4.9 & 0.0148 &  \\
 1.73 & 0.44 & 7.2 & 14 & - & \textit{0.0287} & \\
 2.17 & 0.54 & 9 & 9 & 5.65 & 0.0165 &  \\
 2.65 & 0.65  & 11 & 11  & 6.43 & 0.0182 & \\
 3.37 & 0.84 &  14 & 14 &  7.54 & 0.0206 & \\

\end{tabular}
\end{center}

\caption{Characteristics of the 2DEG sample: electron density $n$,
low temperature mobility $\mu$, perpendicular magnetic field
$B_{F}$ corresponding to $\nu=n/(eB_{F}/h)=1$, total magnetic
field $B_{tot}$, effective Coulomb energy to disorder ratio
$\gamma_{int}$, Zeeman to Coulomb energy ratio $\eta$ (see text
for definitions).} \label{table1}
\end{table}
The $^{71}Ga$ resonance was measured in a dilution fridge inserted
in a 16T superconducting magnet using the recently developed
``frequency-pulsed Resistively-detected'' NMR (f-PRDNMR) technique
\cite{Stern2012,SM.RDNMR}. This technique, a variation of the
``power-pulsed'' resistive NMR \cite{Kumada2007,Tiemann2012},
overcomes the limitation of standard ``continuous wave'' resistive
NMR \cite{Desrat2002} and has recently allowed the study of
various many-body phases in the QH regime
\cite{Stern2012,Tiemann2014,Friess2014}. In the present case, the
NMR detection point where the longitudinal resistance $R_{xx}$ was
measured was chosen in the flank of the $\nu=1$ QH state
(typically $\nu=0.8-0.9$). A large delay time between each filling
factor acquisition was allowed for to make sure the nuclear system
had completely relaxed, and slow sweeps were employed to approach
a static response. The electronic temperature was systematically
determined from the resistance of the sample and its calibrated
temperature dependence, and for the reported scans was essentially
current-limited to a value of about 0.2 K (currents of about 50 nA
were necessary to achieve a sufficient signal-to-noise-ratio to
study the weak response close to $\nu=1$).

\begin{figure}[!h]
\begin{center}
\includegraphics[width= 8.5cm]{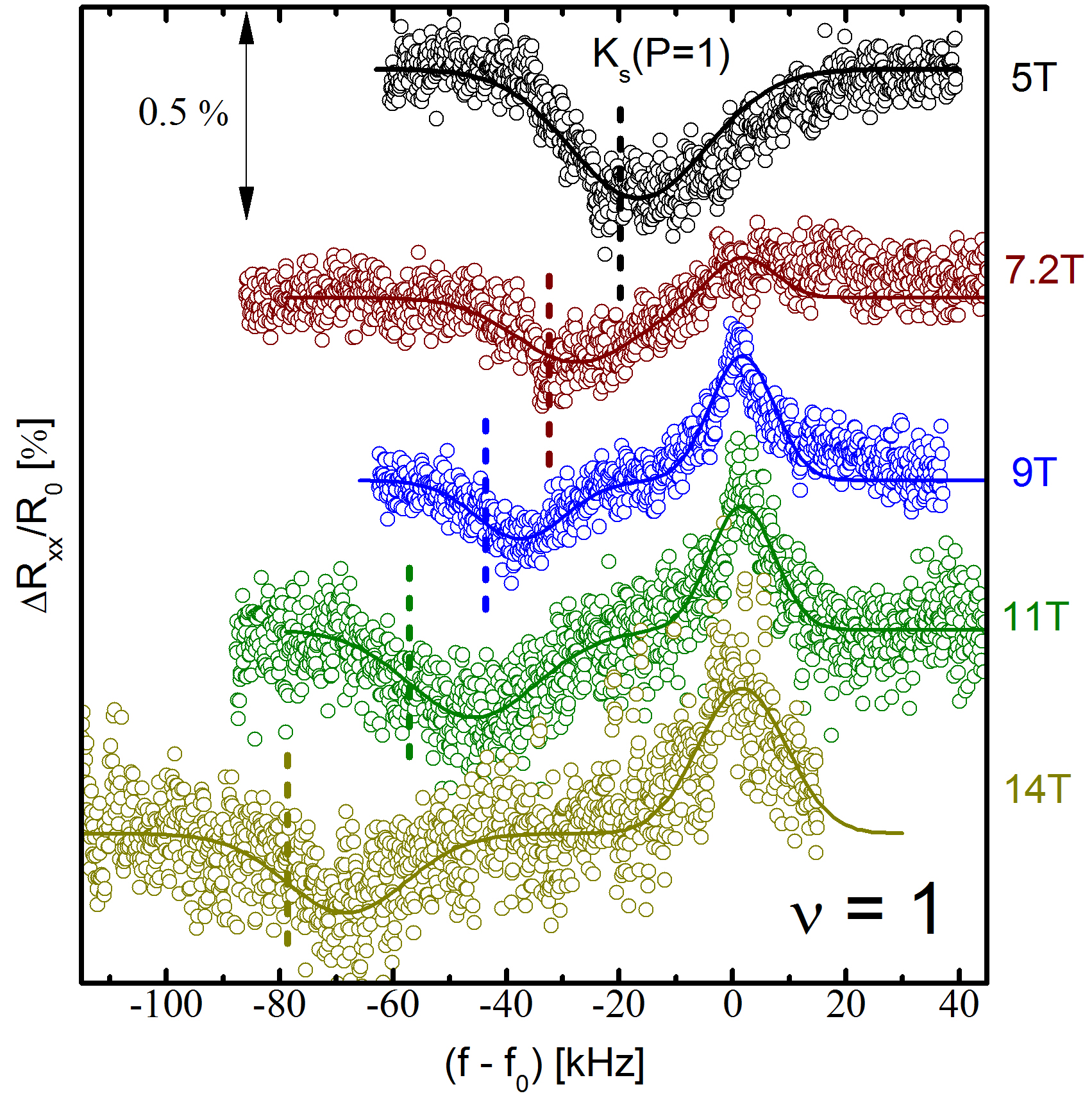}
\end{center}
\caption{(color online)f-PRDNMR signals at exact filling factor
$\nu=$1.000(2) for different magnetic fields (circles). $B_{F}=5$T
(black), 7.2T, 9T, 11T, and 14T. $f_{0}$ is the reference
frequency corresponding to an unpolarized electron system (no
Knight shift). The vertical dashed line denote the Knight shift
for a fully spin polarized system (see text). The solid lines are
simulations of the NMR response \cite{SM.RDNMR}. Curves are offset
vertically for clarity. }\label{Fig1}
\end{figure}

In figure \ref{Fig1}, we report f-PRDNMR signals in a region
previously unexplored by conventional resistive NMR
\cite{Desrat2013,Desrat2002}, namely deep in the QH state where
the longitudinal resistance vanishes. The variation of the
longitudinal resistance $\Delta R_{xx}$ with respect to its
off-resonant value $R_{0}$ is reported as a function of the
relative excitation frequency ($f-f_{0}$), where $f_{0}$ is the
resonance frequency of an unpolarized electron system (zero Knight
shift). The $f_{0}$ value was determined by using the f-PRDNMR
response of unpolarized electron ($P=0$) domains forming at
filling factor $\nu=2/3$ \cite{Stern2004,Stern2012}. Because of
the coexistence of unpolarized and fully polarized domains at this
filling factor, the same measurement could be used to determine
the ``maximal'' value of the Knight shift $\Delta f_{2/3}$, given
by the $P=1$ response observed at lower frequency. This latter
value enables us to locate the expected resonance position for a
fully polarized electron system at $\nu=1$ (vertical dashed-lines
in figure \ref{Fig1}) \cite{SM.Pola}. A peak in the resistive NMR
signal generally sits close to the $P=0$ reference frequency, as
previously observed and discussed in
Refs.\onlinecite{Desrat2013,Desrat2015}. This response, which can
be attributed to an unpolarized electronic sub-system, will not be
discussed here as we focus on the electronic spin polarization
given by the Knight-shifted low frequency response. At low
magnetic field, the resistance minimum is close to the dashed
lines, showing that the polarization at $\nu=1$ is nearly
complete. As the magnetic field $B_{F}$ is increased, the position
of the minimum shifts to the left because the raw Knight shift is
proportional to the electron density. However, the minimum does
not shift as far as the dashed-line, indicating that the
polarization of the system at $\nu=1$ is diminishing as the
magnetic field is increased. This points to a weakening of the QHF
phase in higher magnetic fields (higher electron density). This
behavior, which is the key result of our studies, can also be
evidenced in the vicinity of $\nu=1$ which we will now discuss.

\begin{figure}[h]
\begin{center}
\includegraphics[width=9cm]{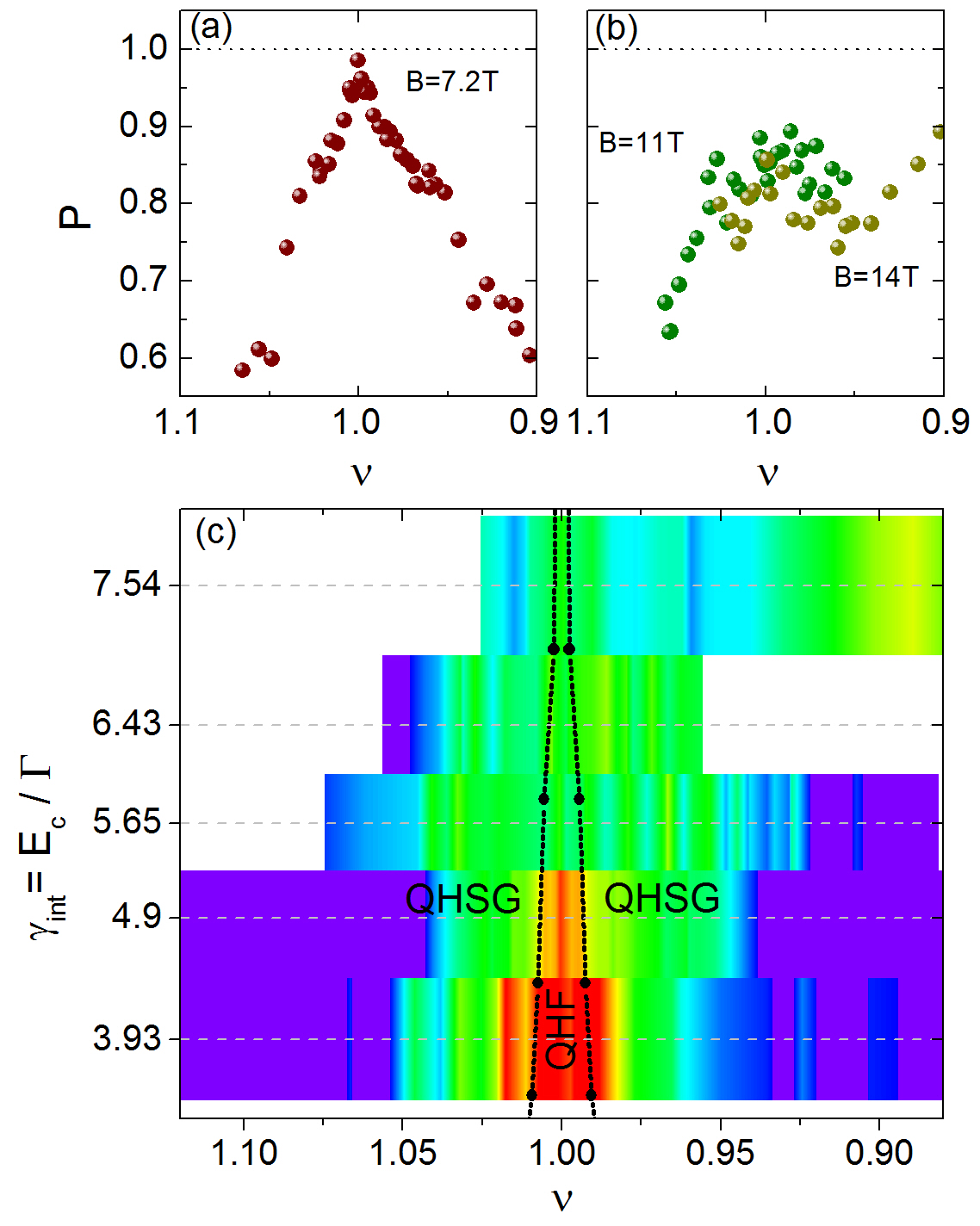}
\end{center}
\caption{(color online) Spin polarization $P$ around filling
factor 1 at magnetic fields of 7.2T (a), 11 and 14T (b). (c)
Polarization color map and phase diagram of the QHF. Color code:
$P\leq0.7$ (purple), $P=0.7$ (blue) to $P=1$(red). The horizontal
axis is the filling factor, and the vertical axis is the
interaction parameter $\gamma_{int}$ (see text). The horizontal
dashed lines represent $\gamma_{int}$ values for which sets of
data were taken. The black dotted line and full circles
materialized the phase boundaries between QHF and QHSG calculated
in reference \cite{Rapsch2002}.}\label{Fig2}
\end{figure}

In figure \ref{Fig2}.a and b we present the extracted filling
factor dependence of the spin polarization for different magnetic
fields. The spin polarization of the system \cite{SM.Pola} at a
filling factor $\nu$ is obtained by:
\begin{equation}\label{Pola}
P_{\nu}=(\Delta f_{\nu} / n_{\nu}) \times (W_{\nu} / \mathcal{A})
\end{equation}
where $\Delta f_{\nu}=f_{r}-f_{0}$ with $f_{r}$ the frequency
corresponding to the $R_{xx}$ minimum, $W_{\nu}$ is the physical
width of the electron system, and $\mathcal{A}$ the hyperfine
coupling constant. The ratio $(W_{\nu}/\mathcal{A})$ is determined
from $\nu=2/3$ calibration experiments, such that the
determination of $P_{\nu}$ is absolute and involves only measured
quantities.

At $B=7.2$T, the polarization of the system tends to be full only
in a narrow filling factor region around $\nu=1$, similarly to
recent observations made in optical absorption experiments
\cite{Plochocka2009}. Away from $\nu=1$, the polarization drops
due to the well-established formation of Skyrmions in the system
\cite{Aifer1996,Barrett1995}. As the magnetic field is increased
(Fig \ref{Fig2}.b), the spin polarization drops and is incomplete
even in the close vicinity of $\nu=1$. In the highest magnetic
fields studied, the average polarization around $\nu=1$ is about
0.75-0.8. The field-induced depolarization globally observed
around $\nu=1$ is not expected from simple a priori
considerations, since in an ideal system the Coulomb interaction
$e^{2}/(4\pi\epsilon_{0}\epsilon_{r}\ell_{B}$)(where the magnetic
length $\ell_{B}=\sqrt{\hbar/eB_{F}}$) should become stronger in
high fields and favor the ferromagnetic state. This is not
expected from Zeeman energy considerations either, since the
increase in the ``Zeeman to Coulomb'' energy ratio
$\eta=(|g^{*}|\mu_{B}B_{tot})/(e^{2}/(4\pi\epsilon_{0}\epsilon_{r}l_{B})$
with the magnetic field should lead to unfavorable conditions for
the skyrmions formation, and thus a repolarization of the system.
This latter effect is actually observed in the 0.85-0.95 filling
factor region for $B=14$T (partially visible in Fig \ref{Fig2}.b),
but absent when going closer to $\nu=1$.

Another important parameter in our experiment is the significant
amount of disorder \cite{SM.dis}. It is well-known that in the
high disorder limit, the $\nu=1$ QH state (and thus the QHF)
collapses \cite{Fogler1995,Sinova2000,Rapsch2002,Pan2011}. While
being larger than in the most recent QHF studies, the disorder in
our sample is still small enough compared to the Coulomb
interaction to ensure that the long range order can develop and
stabilize the QHF. This is theoretically expected for a ratio
$\gamma_{int}$ between Coulomb energy and disorder larger than
$\sim$ 2 \cite{Sinova2000}. On the other hand, a less intuitive
effect of disorder can occur as the 2DEG is slightly taken away
from the $\nu=1$ filling factor. In this situation, charges are
introduced into the system by forming Skyrmions. In the presence
of disorder, fluctuations of the impurity potential generate
random potential wells which establish an optimal skyrmion size
\cite{Nederveen1999,Rapsch2002}. The QHF can resist until the
Skyrmions are numerous enough to overlap, which occurs for a
sufficient deviation $\delta\nu$ from $\nu=1$ determined by the
Skrymion size, and thus, the strength of disorder. These
considerations lead Rapsch \textit{et al.} to build a phase
diagram where the QHF dominates the QH Skyrmion glass (QHSG) over
a small optimal ``window'' of $\gamma_{int}$ and $\delta\nu$
values (figure 2 in Ref.\onlinecite{Rapsch2002}). Increasing
$\gamma_{int}$ at a fixed $\delta\nu$ brings the system to the
Skyrmion glass phase, leading to the peculiar prediction that the
QHF could be destabilized in a more interacting and/or less
disordered system \cite{Sinova2000,Rapsch2002}.

In the following, we show that we are here experimentally probing
this so-far unexplored part of the phase diagram, and the
associated transition between the QHF and the skyrmion glass phase
with increasing $\gamma_{int}$. Figure 2.c. shows a color plot of
the spin polarization, as a function of the filling factor for
different values of interaction ratio $\gamma_{int}$  obtained by
performing experiments at magnetic fields of 5T, 7.2T, 9T and 11T
and 14T. The effective Coulomb energy in our system has been
estimated by performing thermal activation transport experiments
at $\nu=1$, taking into account the contribution of Zeeman energy
and disorder. This enables us to come up with a realistic value of
the Coulomb interaction, about $ 10 \sqrt{B}$ K, taking into
account its large reduction in a non-zero thickness system.
Disorder, more precisely the Landau level full width at half
maximum (FWHM) $\Gamma$, has been estimated by Shubnikov-de Haas
(SdH) measurements. These estimations lead us to an interaction
ratio $\gamma_{int}=(10\sqrt{B})/\Gamma$ varying from about 4 to
7.5 in our experiments \cite{SM.gammaratio}. The theoretical phase
diagram of Ref. \onlinecite{Rapsch2002} is also reported in figure
\ref{Fig2}.c, where the QHF phase and the QHSG phase are separated
by black dotted lines and full circles. We make a quantitative
comparison between our data and theory by defining the ``high
disorder'' phase boundary at $\delta\nu=0$ in Ref.
\onlinecite{Rapsch2002} to match the one in Ref.
\onlinecite{Sinova2000}, which occurs for a value of
$\gamma_{int}=1.6$. In figure \ref{Fig2}, the QHF phase appears as
a ``stripe'' which sharpens with increasing $\gamma_{int}$. As can
be seen, a realistic estimation of $\gamma_{int}$ in our sample
shows that our experiment sits at low magnetic fields in the
``optimal window'' of the phase diagram where the QHF is stable
and the spin polarization complete close to $\nu=1$. Increasing
$\gamma_{int}$ leads to a depolarization in the $\nu=1$ vicinity,
as experimentally observed. These observations also explain why
the QHF is surprisingly more fragile in higher mobility samples,
\cite{Barrett1995,Aifer1996,Plochocka2009,Tiemann2012} where the
higher value of $\gamma_{int}$ confines the QHF to an even
narrower stripe around $\nu=1$.

We note that, even though theories \cite{Sinova2000,Rapsch2002} do
not predict any depolarization in the ideal $\nu=1$ and $T=0$K
cases (the phase boundary tends to $\delta\nu=0$ for infinite
values of $\gamma_{int}$), a depolarization is observed
experimentally at exact $\nu=1$, as already reported in figure
\ref{Fig1}. We attribute this effect to the small but non-zero
inhomogeneity (fluctuations) in the electron density, which makes
the filling factor not perfectly equal to 1 in the physical space,
and consequently pushes the system away from of the $\delta\nu=0$
condition. Even with density inhomogeneities $\delta n/n$ of less
than $\pm0.5 \%$ (an upper bound estimated from low field Hall and
SdH measurements), we are probing at $\nu=1$ a stripe of width
$\delta\nu=\pm0.005$ which extends out of the narrow QHF
theoretical stripe at $\gamma_{int}\sim6$. This is precisely in
the region where we experimentally observe the $\nu=1$
depolarization. An additional loss of spin polarization is induced
by the finite (non-zero) temperature of our experiment. The above
point illustrates the sharpness of the QHF phase, which can only
be stabilized by low temperatures, closeness to $\nu=1$ (implying
high electron density homogeneity), and, as we demonstrated, with
the help of disorder. We recall that a too high amount of disorder
will induce a transition to a QHSG paramagnetic state for
$\gamma_{int} < 1.6$ \cite{Sinova2000,Rapsch2002,Pan2011}, which
defines an optimal amount of disorder to stabilize the QHF.

Finally, we would like to comment on the role of the Zeeman
energy. As we mentioned above, the increase in $\eta$ at higher
perpendicular magnetic-fields (quantified in table \ref{table1})
leads to a repolarization of the system seen in the top right
corner of figure \ref{Fig2}.c. To further enhance the effect of
the Zeeman energy, we have performed tilted field experiments
where the Coulomb energy is limited by the magnetic field
perpendicular to the sample, while the Zeeman energy scales with
the (larger) total magnetic field. This enables us to boost the
value of $\eta$ up to about $\sim 0.03$ (the electronic g-factor
is $g^{*}=-0.44$).
\begin{figure}[!h]
\begin{center}
\includegraphics[width= 8cm]{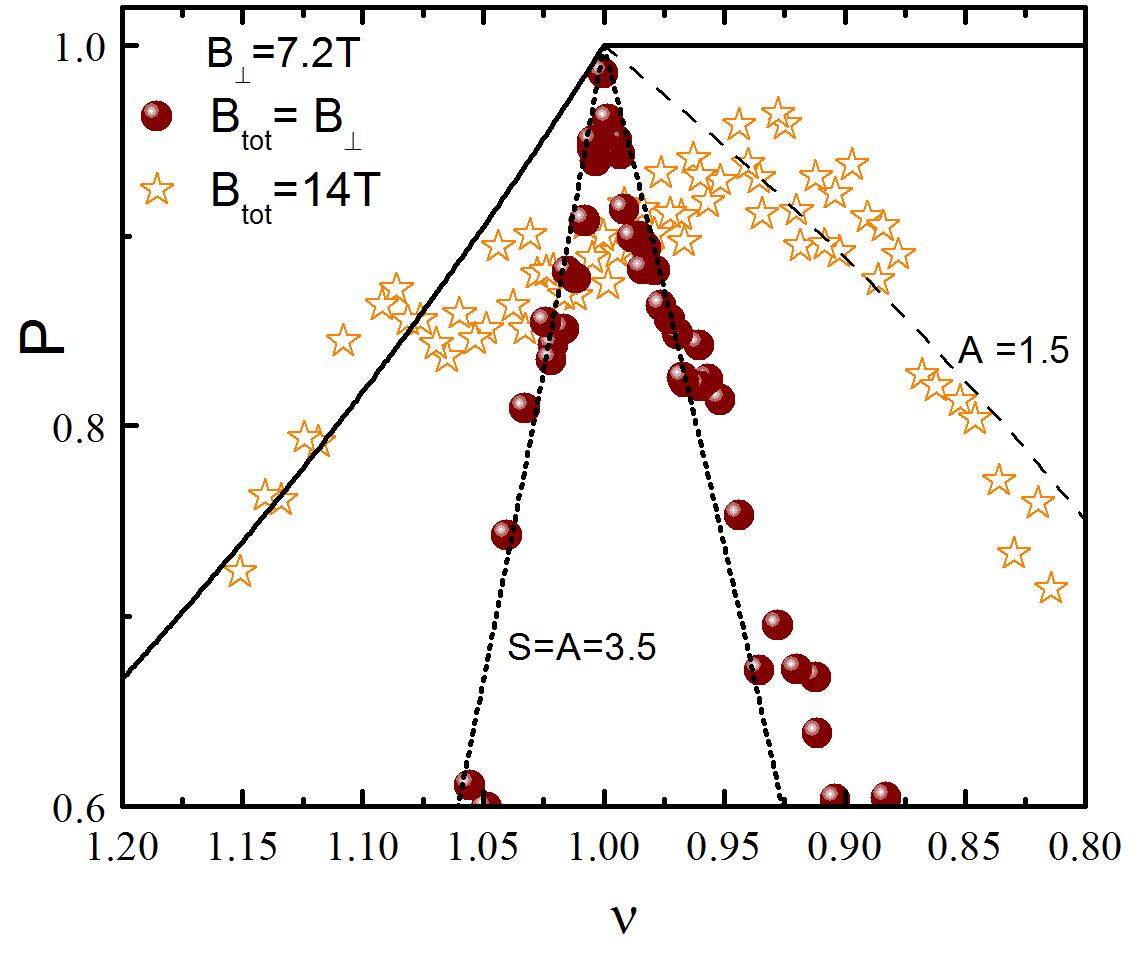}
\end{center}
\caption{(color online) Effect of the Zeeman energy. Spin
polarization $P$ for $\eta=0.0148$ (Dots) and  $\eta=0.0287$
(stars). Dotted (dashed) lines are theoretical expectation for a
macroscopic spin of 3.5 (1.5) per flux quantum, while the solid
lines are the single particle spin polarization.}\label{Fig3}
\end{figure}
In figure \ref{Fig3}, we report the spin polarization around
$\nu=1$ in the two cases $\eta=0.0148$ (data of figure
\ref{Fig2}.a) and $\eta=0.0287$. In the first case, the
depolarization away from $\nu=1$ lies very close to the
theoretical expected behaviour due to the formation of Skyrmions
(anti-Skyrmions) of size S (A)=3.5 for $\eta=0.014$ (dashed lines)
\cite{Sondhi1993}. In the second case, the polarization drop away
from $\nu=1$ is much less pronounced, showing that skyrmions are
destabilized by the large value of $\eta$, again, in agreement
with theoretical predictions. More precisely, for $\nu>1.1$ the
spin polarization follows the single-particle spin polarization
with single spin flip excitation ($S=1$), theoretically expected
for $\eta=0.03$ \cite{Abolfath1997}, very close to our
experimental conditions. Quantitatively recovering the single
particle spin polarization is an additional confirmation of the
accuracy of our absolute polarization measurement. For $\nu<0.95$,
very small size ($A=1.5$) anti-skyrmions are observed.

To conclude, we have reported absolute low temperature spin
polarization measurements of the quantum Hall ferromagnet by
employing state-of-the start resistive NMR technique. Our results
show that an optimal amount of disorder can stabilize the fully
polarized QHF on a narrow temperature/filling factor window by
preventing Skyrmion from expanding in the 2DEG. This accounts for
the surprising fragility of this phase in high quality 2D systems,
and opens up new ways to generate robust 2D quantum Hall
ferromagnets.

\end{document}